\documentclass[12pt]{article}
\usepackage{epsfig}
\def\be{\begin{equation}}
\def\ee{\end{equation}}
\def\bea{\begin{eqnarray}}
\def\eea{\end{eqnarray}}
\usepackage{graphicx}

\catcode`\@=11
\def\lsim{\mathrel{\mathpalette\@versim<}}
\def\gsim{\mathrel{\mathpalette\@versim>}}
\def\@versim#1#2{\vcenter{\offinterlineskip
\ialign{$\m@th#1\hfil##\hfil$\crcr#2\crcr\sim\crcr } }}
\catcode`\@=12

\catcode`@=12
\begin{document}
\thispagestyle{empty}
\begin{flushright}
UCRHEP-T583\\
Aug 2017\
\end{flushright}
\vspace{0.6in}
\begin{center}
{\LARGE \bf Cobimaximal Neutrino Mixing from $A_4$\\ 
and its Possible Deviation\\}
\vspace{1.0in}
{\bf Ernest Ma$^{a,*}$ and G. Rajasekaran$^{b,c,\dagger}$\\}
\vspace{0.2in}
{$^a$ \sl Physics \& Astronomy Department and Graduate Division,\\ 
University of California, Riverside, California 92521, USA\\}
\vspace{0.1in}
{$^b$ \sl The Institute of Mathematical Sciences, Chennai 600 113, India\\}
\vspace{0.1in}
{$^c$ \sl Chennai Mathematical Institute, Chennai 603 103, India\\}
\end{center}
\vspace{1.0in}

\begin{abstract}\
It has recently been shown that the phenomenologically successful pattern of 
cobimaximal neutrino mixing ($\theta_{13} \neq 0$, $\theta_{23} = \pi/4$, 
and $\delta_{CP} = \pm \pi/2$) may be achieved in the context of 
the non-Abelian discrete symmetry $A_4$, if the neutrino mass matrix 
is diagonaized by an orthogonal matrix ${\cal O}$.  We study how this 
pattern would deviate if ${\cal O}$ is replaced by an unitary matrix.  
\end{abstract}
\vspace{0.4in}

$^*$email: ma@phyun8.ucr.edu

$^\dagger$email: graj@imsc.res.in

\newpage
\baselineskip 24pt

\noindent \underline{Introduction}:

Present neutrino data~\cite{pdg2016,t2k15} are suggestive of the pattern 
of  \underline{cobimaximal} mixing~\cite{m15}, i.e. $\theta_{13} \neq 0$, 
$\theta_{23} = \pi/4$ and $\delta_{CP} = -\pi/2$, which was discussed 
many years ago ahead of the data.  It was first derived~\cite{fmty00,mty01} 
using the ansatz
\begin{equation}
U_{l \nu} = U_\omega {\cal O},
\end{equation}
where~\cite{c78,w78}
\begin{equation}
U_\omega = {1 \over \sqrt{3}} \pmatrix{1 & 1 & 1 \cr 1 & \omega & \omega^2 \cr 
1 & \omega^2 & \omega}
\end{equation}
with $\omega = \exp(2 \pi i/3) = -1/2 + i \sqrt{3}/2$, and ${\cal O}$ is 
any arbitrary real orthogonal matrix.  This intriguing idea lay dormant 
for many years until two years ago when it was shown~\cite{m15-1,h15} how 
$A_4$~\cite{mr01} may be used to realized this possibility.  The origin 
of ${\cal O}$ comes from the scotogenic generation of neutrino mass from 
a set of real scalars~\cite{m15-1,fmp14,mnp15,fmz16,m16}

There is a second equivalent way to cobimaximal mixing which deals directly 
with the $3 \times 3$ Majorana neutrino mixing matrix in the basis where 
charged-lepton masses are diagonal.  It was discovered also in the context 
of $A_4$, i.e. it should be of the form~\cite{m02,bmv03,gl04}
\begin{equation}
{\cal M}_\nu^{(e,\mu,\tau)} = \pmatrix{A & C & C^* \cr C & D^* & B \cr 
C^* & B & D},
\end{equation}
where $A,B$ are real.  In either case, the conditions 
$|U_{\mu i}| = |U_{\tau i}|$ for $i=1,2,3$ are obtained, which lead to 
cobimaximal mixing.  This conceptual shift from 
tribimaximal mixing~\cite{hps02,m04}, i.e. 
$\theta_{13}=0, \theta_{23}=\pi/4, \tan^2 \theta_{12} = 1/2$, to 
cobimaximal mixing may also be understood 
as the result of a residual generalized $CP$ 
symmetry~\cite{gl04,cyd15,mn15,jp15,hrx15,n16,lld16}.  

\noindent \underline{$S_3 \times Z_2$ realization of cobimaximal mixing}:

Whereas Eq.~(1) has been used in the context of $A_4$ to obtain cobimaximal 
neutrino mixing, it has also been shown very recently~\cite{m17} 
that it may be 
accomplished with $U_\omega$ replaced by
\begin{equation}
U_2 = \pmatrix{1 & 0 & 0 \cr 0 & 1/\sqrt{2} & -i/\sqrt{2} \cr 0 & 1/\sqrt{2} 
& i/\sqrt{2}},
\end{equation}
which is derivable from $S_3 \times Z_2$.  In that case, if ${\cal O}$ is 
replaced by
\begin{equation}
U = \pmatrix{1 & 0 & 0 \cr 0 & c_{23} & s_{23} \cr 0 & -s_{23} & c_{23}} 
\pmatrix{c_{13} & 0 & s_{13}e^{-i \delta} \cr 0 & 1 & 0 \cr -s_{13} 
e^{i \delta} & 0 & c_{13}} \pmatrix{c_{12} & s_{12} & 0 \cr -s_{12} & c_{12} 
& 0 \cr 0 & 0 & 1},
\end{equation}
where $c_{ij} = \cos \phi_{ij}$ and $s_{ij} = \sin \phi_{ij}$, multiplied 
on the right by a diagonal Majorana phase matrix
\begin{equation}
\pmatrix{1 & 0 & 0 \cr 0 & e^{i \alpha_{21}/2} & 0 \cr 
0 & 0 & e^{i \alpha_{31}/2}},
\end{equation}
then the deviation from cobimaximal mixing if $\delta \neq 0$ is very simple. 
The resulting mixing matrix has $\theta_{23} = \pi/4$ regardless of 
$\delta_{CP} = -\pi/2 + \delta$.  This comes from
\begin{equation}
U_2 \pmatrix{1 & 0 & 0 \cr 0 & c_{23} & s_{23} \cr 0 & -s_{23} & c_{23}} 
= \pmatrix{1 & 0 & 0 \cr 0 & e^{i \phi_{23}} & 0 \cr 0 & 0 & 
-e^{-i \phi_{23}}} \pmatrix{1 & 0 & 0 \cr 0 & 1/\sqrt{2} & -i/\sqrt{2} \cr 
0 & -1/\sqrt{2} & -i/\sqrt{2}}.
\end{equation}
The diagonal matrix of phases on the left may then be absorbed into the 
charged 
leptons, and the remaining part of $U_2 U$ becomes 
\begin{eqnarray}
U_\delta &=& \pmatrix{1 & 0 & 0 \cr 0 & 1/\sqrt{2} & 1/\sqrt{2} \cr 0 & 
-1/\sqrt{2} & 1/\sqrt{2}} 
\pmatrix{c_{13} & 0 & is_{13}e^{-i \delta} \cr 0 & 1 & 0 \cr is_{13} 
e^{i \delta} & 0 & c_{13}} \pmatrix{c_{12} & s_{12} & 0 \cr -s_{12} & c_{12} 
& 0 \cr 0 & 0 & 1} \nonumber \\ 
&=& \pmatrix{c_{12} c_{13} & s_{12} c_{13} & is_{13} e^{-i\delta} \cr 
(-s_{12} + ic_{12} s_{13} e^{i \delta})/\sqrt{2} & (c_{12} + i s_{12} s_{13} 
e^{i \delta})/\sqrt{2} & c_{13}/\sqrt{2} \cr  
(s_{12} + ic_{12} s_{13} e^{i \delta})/\sqrt{2} & (-c_{12} + i s_{12} s_{13} 
e^{i \delta})/\sqrt{2} & c_{13}/\sqrt{2}}  
\end{eqnarray}
multiplied on the right by the diagonal phase matrix
\begin{equation}
\pmatrix{1 & 0 & 0 \cr 0 & 1 & 0 \cr 
0 & 0 & -i}.
\end{equation}
This shows that if $\delta=0$, cobimaximal mixing is achieved with 
$e^{-i \delta_{CP}} = e^{i\pi/2} = i$ as expected.  However, even if 
$\delta \neq 0$, so that $\delta_{CP}$ deviates from $-\pi/2$,  
$\theta_{23}$ remains at $\pi/4$.  Note also that the input $\phi_{12}$ and 
$\phi_{13}$ angles are also the output $\theta_{12}$ and $\theta_{13}$ angles. 
This is a remarkable result and it is only 
true because of $U_2$, and does not hold for $U_\omega$.

\noindent \underline{$A_4$ realization of cobimaximal mixing}:

To study how $U_\omega$ would change the above result, we consider
\begin{equation}
U_{l \nu} = U_\omega U.
\end{equation}
The first thing we note is that if $\delta = 0$ in $U$, then Eq.~(10) 
becomes Eq.~(1) and we obtain cobimaximal mixing.  To find the deviation 
for $\delta \neq 0$, we use
\begin{eqnarray}
U_{e2} &=& {1 \over \sqrt{3}}[s_{12}c_{13}+c_{12}(c_{23}-s_{23})-s_{12}(s_{23}+c_{23})
s_{13}e^{i\delta}], \\
U_{e3} &=& {1 \over \sqrt{3}}[s_{13}e^{-i\delta} + c_{13}(s_{23}+c_{23})], \\
U_{\mu2} &=& {1 \over \sqrt{3}}[s_{12}c_{13}+c_{12}(\omega c_{23}-\omega^2 s_{23}) 
-s_{12}s_{13}e^{i\delta}(\omega s_{23} + \omega^2 c_{23})], \\ 
U_{\mu 3} &=& {1 \over \sqrt{3}}[s_{13}e^{-i\delta} + c_{13} (\omega s_{23} + 
\omega^2 c_{23})],
\end{eqnarray}
then
\begin{eqnarray}
\sin^2 \theta_{13} &=& |U_{e 3}|^2 = {1 \over 3}[s_{13}\cos \delta + c_{13} 
(c_{23} + s_{23})]^2 + {1 \over 3} s_{13}^2 \sin^2 \delta \nonumber \\ 
&=& {1 \over 3} [1 + 2c_{13}^2 s_{23} c_{23} + 2s_{13}c_{13}(c_{23}+s_{23})
\cos \delta], \\ 
\sin^2 \theta_{23} &=& {|U_{\mu 3}|^2 \over 1-|U_{e 3}|^2} = {1 \over 2} + 
{s_{13}c_{13}(c_{23}-s_{23})\sin \delta \over \sqrt{3}(1-\sin^2 \theta_{13})}, \\
\sin^2 \theta_{12} &=& {|U_{e2}|^2 \over 1-|U_{e 3}|^2} \\ 
&=& {|s_{12}c_{13} + c_{12}
(c_{23}-s_{23}) - (c_{23}+s_{23})s_{12}s_{13} \cos \delta|^2 + 
(c_{23} + s_{23})^2s^2_{12}s^2_{13} \sin^2 \delta \over 3(1-\sin^2 \theta_{13})}, 
\nonumber 
\end{eqnarray}
and
\begin{eqnarray}
\cos \delta_{CP} &=& {1 - \sin^2 \theta_{23} - \sin^2 \theta_{12} (1 - \sin^2 
\theta_{23} (1+\sin^2 \theta_{13}))- |U_{\mu2}|^2 \over 2 \sin \theta_{13} 
\sin \theta_{12} \cos \theta_{12} \sin \theta_{23} \cos \theta_{23}} 
\nonumber \\ &=& {-s_{13} \sin \delta \over 2 \sqrt{3} \sin \theta_{13} 
\sin \theta_{12} \cos \theta_{12} \sin \theta_{23} \cos \theta_{23}} \\
&\times& \left[ {c_{13}(c_{23}-s_{23})[1-\sin^2 \theta_{12}(1+\sin^2 
\theta_{13})] \over (1-\sin^2 \theta_{13})} + s_{12}[c_{12}-s_{12}c_{13}
(c_{23}-s_{23})] \right]. \nonumber
\end{eqnarray}

\newpage
\noindent \underline{Deviations from cobimaximal mixing}:

In Eqs.~(16) and (18), if $\delta = 0$, then $\theta_{23} = \pi/4$ and 
$\cos \delta_{CP} = 0$ as expected.  The deviations from these values 
depend not only on $\delta \neq 0$, but also on the input values of 
$s_{ij},c_{ij}$.  They are of course constrained by the output values 
which must agree with data~\cite{pdg2016}, i.e.
\begin{eqnarray}
\sin^2 \theta_{13} &=& 0.021 \pm 0.0011, \\ 
\sin^2 \theta_{12} &=& 0.307 \pm 0.013, \\ 
\sin^2 \theta_{23} &=& 0.51 \pm 0.04 ~(\rm normal~ordering), \\ 
\sin^2 \theta_{23} &=& 0.50 \pm 0.04 ~(\rm inverted~ordering). 
\end{eqnarray}
We first discuss some special cases which keep $\theta_{23} = \pi/4$. 
\begin{itemize}
\item{$s_{13}=0$: In this case, $\delta$ disappears from $U$, so Eq.~(10) 
becomes Eq.~(1) and cobimaximal mixing is assured.  Using Eq.~(15), we 
obtain $2s_{23}c_{23} = -0.937$, resulting in the solutions:
\begin{eqnarray}
&& c_{23} = 0.82137, ~~ s_{23} = -0.57040, \\
&& c_{23} = 0.57040, ~~ s_{23} = -0.82137.
\end{eqnarray}
In either case, $c_{23}-s_{23} = 1.39177$.  Using Eq.~(17), we then obtain 
\begin{eqnarray}
&& c_{12} = 0.93572, ~~ s_{12} = -0.35274, \\ 
&& c_{12} = 0.03582, ~~ s_{12} = -0.99936.
\end{eqnarray}

Note that the arbitrary choices of $c_{12} = 0$ and $s_{23} = -1/\sqrt{3}$ 
would result in 
\begin{eqnarray}
&& \sin \theta_{13} = {\sqrt{2}-1\over 3}, ~~~ \sin \theta_{12} = 
{1 \over \sqrt{3} \cos \theta_{13}}, ~~~ \cos^2 \theta_{13} = 
{2 \over 9}(3+\sqrt{2}), \\ && \sin \theta_{12} \cos \theta_{13} = 
{1 \over \sqrt{3}}, ~~~ \cos \theta_{12} \cos \theta_{13} = 
{\sqrt{2}+1 \over 3}.
\end{eqnarray}
Hence
\begin{equation}
\sin^2 \theta_{12} = {3(3-\sqrt{2}) \over 14} = 0.34, ~~~  
\sin^2 \theta_{13} = {3 -2\sqrt{2} \over 9} = 0.019, 
\end{equation}
which are within 2.5$\sigma$ and 1.8$\sigma$ of the data respectively.  
The neutrino mixing matrix is then given by
\begin{equation}
U_{l\nu} = \pmatrix{-(\sqrt{2}+1)/3 & 1/\sqrt{3} & (\sqrt{2}-1)/3 \cr 
(\sqrt{2}+1)/6-i(\sqrt{2}-1)/2\sqrt{3} & 1/\sqrt{3} & 
-(\sqrt{2}-1)/6-i(\sqrt{2}+1)/2\sqrt{3} \cr 
(\sqrt{2}+1)/6+i(\sqrt{2}-1)/2\sqrt{3} & 1/\sqrt{3} & 
-(\sqrt{2}-1)/6+i(\sqrt{2}+1)/2\sqrt{3}},
\end{equation}
and the neutrino mass matrix is of the form
\begin{equation}
{\cal M}_\nu = \pmatrix{m_2 & 0 & 0 \cr 0 & (2m_1+m_3)/3 & \sqrt{2}(m_1-m_3)/3 
\cr 0 & \sqrt{2}(m_1-m_3)/3 & (m_1+2m_3)/3}.
\end{equation}
Using the invariant
\begin{eqnarray}
J_{CP} &=& Im(U_{\mu 3} U^*_{e3} U_{e2} U^*_{\mu 2}) = {-1 \over 18 \sqrt{3}} 
\\ &=& \sin \theta_{13} \cos \theta_{13}^2 \sin \theta_{12} \cos \theta_{12} 
\sin \theta_{23} \cos \theta_{23} \sin \delta_{CP}, \nonumber
\end{eqnarray}
we obtain $\delta_{CP} = -\pi/2$ as expected.
}
\item{$c_{13}=0$:  Whereas this implies $\sin^2 \theta_{23} = 1/2$, it also 
requires $\sin^2 \theta_{13}= 1/3$ which is of course not supported by the 
data.}
\item{$c_{23} - s_{23}=0$:  In this case, $c_{23} = s_{23} = 1/\sqrt{2}$, 
and even if $\delta \neq 0$, $\sin^2 \theta_{23} = 1/2$ from Eq.~(16). 
Now
\begin{eqnarray}
\sin^2 \theta_{13} &=& {1 \over 3} [1 + c_{13}^2 + 2 \sqrt{2} s_{13} c_{13} 
\cos \delta], \\ 
\sin^2 \theta_{12} &=& {s_{12}^2[2-c_{13}^2 - 2 \sqrt{2} s_{13} c_{13} 
\cos \delta] \over 3(1-\sin^2 \theta_{13})} = s_{12}^2, \\ 
\cos \delta_{CP} &=& {- s_{13} \sin \delta \over \sqrt{3} \sin \theta_{13}}.
\end{eqnarray}
If $\delta = 0$, then $\cos \delta_{CP} = 0$ and 
\begin{equation}
s_{13}^2 = {1 \over 3} [2-\sin^2 \theta_{13} \pm 2 \sqrt{2} \sin \theta_{13} 
\cos \theta_{13}].
\end{equation}
If $\delta \neq 0$, then
\begin{equation}
s_{13}^2 = {1.937 + 4\cos^2 \delta \pm 4 \cos \delta \sqrt{\cos^2 \delta 
- 0.9075} \over 1 + 8\cos^2 \delta}.
\end{equation}
Plugging this into Eq.~(35), we then obtain $\cos \delta_{CP}$ as 
a function of $\delta$.  As $\cos^2 \delta$ ranges from 1 to 0.9075, 
$\cos \delta_{CP}$ goes from 0 to $\pm 0.995$.  This means that almost 
any value of $\delta_{CP}$ is allowed.
}
\end{itemize}
We now consider a special case with $\theta_{23} \neq \pi/4$.
\begin{itemize}
\item{$c_{23} + s_{23}=0$:  In this case, $c_{23} = 1/\sqrt{2}$, 
$s_{23} = -1/\sqrt{2}$.  Hence $s_{13} = \sqrt{3} \sin \theta_{13}$, and 
$s_{12}$ is expressible in terms of $\sin \theta_{12}$ and $\sin \theta_{13}$. 
The deviation of $\sin^2 \theta_{23}$ from 1/2 and that of $\cos \delta_{CP}$ 
from zero are now correlated.  For $\sin^2 \theta_{13} = 0.021$ and 
$\sin^2 \theta_{12}=0.307$, we obtain $s^2_{12} = 0.9998$ or 0.1399. 
For $s^2_{12} = 0.9998$, we find
\begin{equation}
\cos \delta_{CP} = 4.223 \left( \sin^2 \theta_{23} - {1 \over 2} \right).
\end{equation}
Allowing $\sin^2 \theta_{23}$ to differ from 1/2 by 0.04, we find that 
$|\delta_{CP}| > 80^\circ$.  
For $s^2_{12} = 0.1399$, we find
\begin{equation}
\cos \delta_{CP} = -5.505 \left( \sin^2 \theta_{23} - {1 \over 2} \right),
\end{equation}
so that $|\delta_{CP}| > 77^\circ$.

If we choose $s^2_{12}=1$ exactly, then the phase $\delta$ may be rotated 
away, so that $U$ becomes ${\cal O}$ and cobimaximal mixing is recovered. 
In addition we have $|U_{e1}| =\sqrt{2/3}$ resulting in
\begin{equation}
\sin^2 \theta_{12} = {1 - 3 \sin^2 \theta_{13} \over 3(1-\sin^2 \theta_{13})} 
= 0.319,
\end{equation}
which is within 1$\sigma$ of the data.  This prediction is equivalent to 
$\tan^2 \theta_{12} = (1 - 3\sin^2 \theta_{13})/2$, which was derived 
previously~\cite{m12} in a completely different way involving the 
conditions 
$\nu_1 = (2 \nu_e - \nu_\mu - \nu_\tau)/\sqrt{6}$ and $\cos \delta_{CP} =0$.  
In our specific case, the neutrino mass matrix ${\cal M}_\nu$ is of the form
\begin{equation}
\pmatrix{c_{13}^2 m_2 + s_{13}^2 m_3 & 
s_{13}c_{13}(m_2-m_3)/\sqrt{2} & -s_{13}c_{13}(m_2-m_3)/\sqrt{2} \cr 
s_{13}c_{13}(m_2-m_3)/\sqrt{2} & (m_1 + s_{13}^2 m_2 + c_{13}^2 m_3)/2 & 
(m_1 - s_{13}^2 m_2 - c_{13}^2 m_3)/2 \cr -s_{13}c_{13}(m_2-m_3)/\sqrt{2} &  
(m_1 - s_{13}^2 m_2 - c_{13}^2 m_3)/2 & (m_1 + s_{13}^2 m_2 + c_{13}^2 m_3)/2}.
\end{equation}
Assuming $m_{1,2,3}$ to be real, this is realized with 4 real parameters 
$A,B,C,D$, i.e.
\begin{equation}
{\cal M}_\nu = \pmatrix{A & D & -D \cr D & B & C \cr -D & C & B}.
\end{equation}
If $D=0$, then it is well-known~\cite{m04} that we obtain tribimaximal mixing. 
With $D \neq 0$, it is also known~\cite{m04} that the above pattern predicts 
$\sin^2 \theta_{12} < 1/3$.  With Eq.~(41), we now have cobimaximal 
mixing as well as the prediction of Eq.~(40).
}
\end{itemize}

\noindent \underline{Concluding remarks}:
Using $U_\omega$ of Eq.~(2) which is derivable from $A_4$, we expand on the 
intriguing result of Eq.~(1), namely that if the $3 \times 3$ Majorana 
neutrino mass matrix is diagonalized by an orthogonal matrix ${\cal O}$, 
then neutrino cobimaximal mixing 
($\theta_{13} \neq 0, \theta_{23}=\pi/4,\delta_{CP}=\pm\pi/2$) 
is guaranteed.  We consider the most general unitary matrix $U$ of Eq.~(5) 
instead of ${\cal O}$ and discuss how cobimaximal mixing is affected in a 
number of simple cases.  As a byproduct, we obtain Eqs.~(31) and (42) 
which are suggestive of underlying patterns of the neutrino mass matrix yet 
to be explored theoretically.  Whereas cobimaximal mixing is a good 
representation of the present data, future more precise measurements 
will be able to distinguish among these different possible scenarios.

\noindent \underline{Acknowledgment}:
This work is supported in part 
by the U.~S.~Department of Energy under Grant No.~DE-SC0008541.

\bibliographystyle{unsrt}

\end{document}